# How accurate is an Arduino Ohmmeter?

Gergely Makan, Robert Mingesz and Zoltan Gingl

Department of Technical Informatics, University of Szeged, Árpád tér 2, 6720, Szeged, Hungary

## Abstract

The Arduino platform is widely used in education of physics to perform a number of different measurements. Teachers and students can build their own instruments using various sensors, the analogue-to-digital converter of the Arduino board and code to calculate and display the result. In several cases this can mean incautious reproduction of what can be found on the Internet and an in-depth understanding can be missing. Here we thoroughly analyse a frequently used resistance measurement method and show demonstration experiments as well to make it clear.

## Construction of the Ohmmeter

A wide range of experiments and solutions are shown where sensors and the Arduino platform are used to teach physics efficiently and attractively while the cost is kept very low [1-5]. Sensors have an output (voltage, current, resistance, capacitance, inductance) that can be handled by electronics. Since an analogue-to-digital converter (A/D converter, ADC) has voltage input, some circuitry is needed to convert the sensor's output into voltage that fits in the input range of the ADC. Like in any measurement, a reference is required. The Arduino board's ADC accepts an input voltage in the range from 0 V to $V_{REF}$, where $V_{REF}$ is the reference voltage. Digital sensors incorporate all of these plus a digital data interface.

The simplest way to measure resistance is to build a voltage divider which has an output voltage that depends on the unknown resistance, see figure 1. This is the common method applied in Arduino Ohmmeter projects published on the Internet.

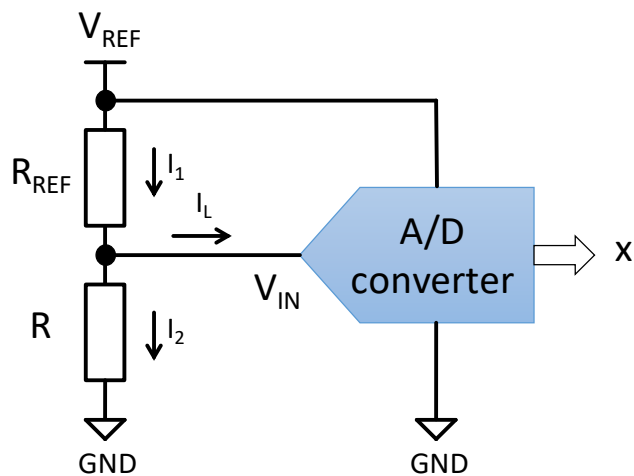

**Figure 1.** A voltage divider can be used to measure resistance. $V_{REF}$ and $R_{REF}$ are known, $R$ can be determined as a function of $x$.

Although some leakage current $I_L$ flows into the input of the ADC, we should use our device in conditions when its effect is negligible, just like in most cases of instrumentation [6].

$V_{REF}$ and $R_{REF}$ are known, therefore the input voltage $V_{IN}$ of the ADC can be easily calculated:

$$V_{IN} = \frac{R}{R + R_{REF}} V_{REF} \tag{1}$$

The output code of the ADC can have *N* different values, for the 10-bit ADC of the Arduino board *N* equals $2^{10}$ = 1024. Therefore, the input voltage can be expressed as:

$$V_{IN} = \frac{x}{N} V_{REF} \tag{2}$$

where the digital output code *x* is an integer in the range from 0 to *N*-1. Note that *N*-1 is frequently used in the Arduino community as the denominator in equation (2), although that is incorrect [7].

The unknown resistance can be given as a function of *x*:

$$R = \frac{R_{REF} \cdot x}{N - x} \tag{3}$$

This highlights the main advantage of the so-called ratiometric arrangement shown in figure 1: *R* does not depend on $V_{REF}$. Therefore, the accuracy of $V_{REF}$ does not matter, the 5 V supply is sufficient even if its accuracy is moderate. Despite this, it is a good practice in any measurement application to externally power the Arduino board instead of using the USB power. Not only because of accuracy, but to avoid excessive noise.

Since *x* is an integer, quantisation error occurs. In addition, since the ADC is not ideal, there are other error sources too, as discussed in the datasheet [8]. The *typical* absolute accuracy is specified as 2 units, so we can assume that the overall error is Δ*x*=±2.

How much error does it mean in the resistance measurement? In order to determine this, first take the derivative of *R* as a function of *x*:

$$\frac{\partial R}{\partial x} = \frac{R_{REF} \cdot N}{(N - x)^2} \tag{4}$$

Considering small deviations, the error of the resistance can be approximated as:

$$|\Delta R_x| \approx \left|\frac{\partial R}{\partial x} \Delta x\right| = \left|\frac{R_{REF} \cdot N}{(N - x)^2} \Delta x\right| \tag{5}$$

and the relative error is obtained by the following expression:

$$\left|\frac{\Delta R_x}{R}\right| \approx \left|\frac{N}{x \cdot (N - x)} \Delta x\right| \tag{6}$$

The value of the reference resistor $R_{REF}$ has some error too. Following the same method, it is easy to calculate its contribution:

$$\frac{\partial R}{\partial R_{REF}} = \frac{x}{N - x} \tag{7}$$

$$|\Delta R_r| \approx \left|\frac{\partial R}{\partial R_{REF}} \Delta R_{REF}\right| = \left|\frac{x}{N - x} \Delta R_{REF}\right| \tag{8}$$

$$\left|\frac{\Delta R_r}{R}\right| \approx \left|\frac{\Delta R_{REF}}{R_{REF}}\right| \tag{9}$$

Therefore, the overall relative error of the unknown resistance can be estimated as:

$$\left|\frac{\Delta R}{R}\right| \approx \left|\frac{\Delta R_x}{R}\right| + \left|\frac{\Delta R_r}{R}\right| \approx \left|\frac{N}{x \cdot (N-x)}\Delta x\right| + \left|\frac{\Delta R_{REF}}{R_{REF}}\right| \qquad (10)$$

Figure 2 shows this relative error as a function of the ADC code *x* for 0.1 %, 1 % reference resistor tolerances and ADC error of Δ*x*=±2.

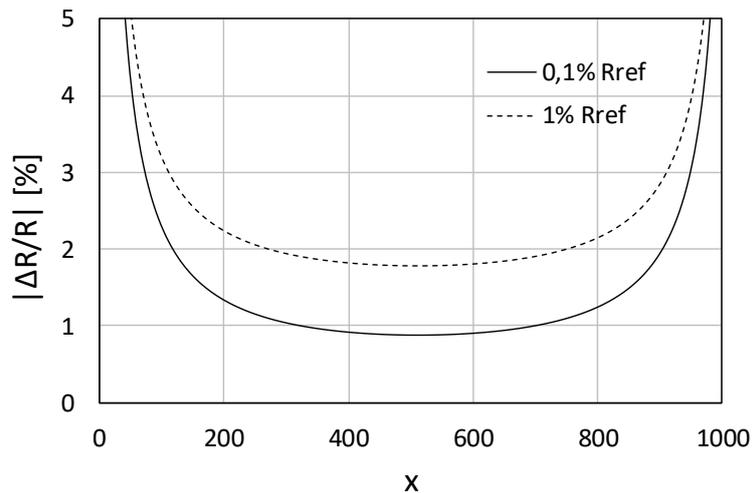

**Figure 2.** Calculated relative measurement error as the function of the ADC code.

## To calibrate or not to calibrate?

Calibration can be harder than expected. Digital multimeters (DMMs) can have an error above 1 % in resistance measurement mode, while you can buy dozens of resistors with 1 % tolerance for cents. More precise DMMs are expensive, they must be regularly calibrated (typically once per two years) at a cost higher than the price of a common DMM.

Simple two-point calibration can eliminate offset and gain errors, but not nonlinearity. Therefore, the absolute error can only be removed by calibrating the ADC for each possible code, which is very complicated and unreasonable.

In conclusion, it is advisable to use a reference resistor with specified tolerance (e.g. 1 % or 0,1 %) and estimate the error as we have shown above. Note that resistance accuracy depends on temperature and some other conditions [9]. Of course, it is important the check if your instrument gives consistent data. Test your Ohmmeter with some known value resistors with sufficient accuracy.

## Testing the Ohmmeter

We have tested the performance of the Ohmmeter and compared the theoretical and measured errors. First, we have measured the resistance of eight resistors to 0,01 % accuracy using a professional instrument (Agilent 34410A). Then we have built four Ohmmeters with Arduino boards from different manufacturers, see figure 3.

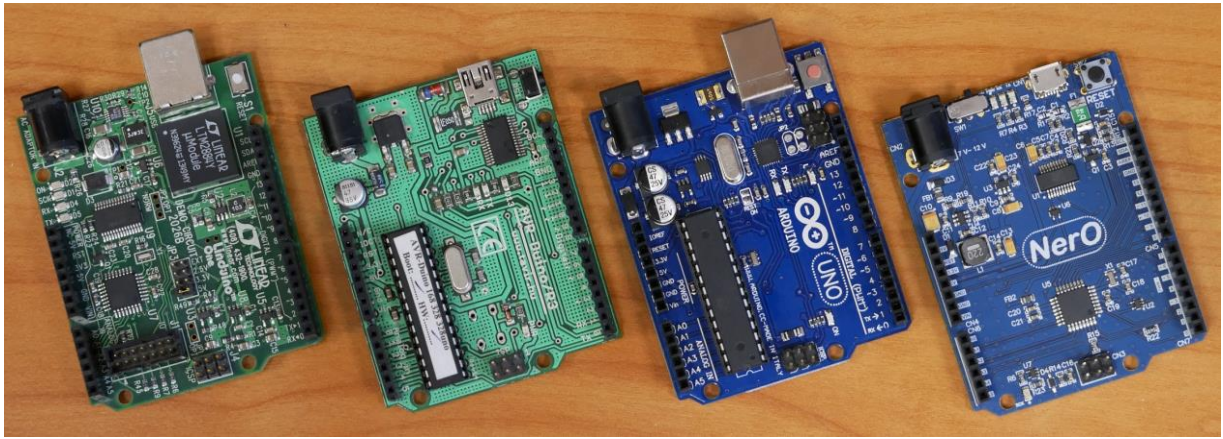

**Figure 3.** Four Arduino-compatible boards from different manufacturers used to build the Ohmmeters.

Figure 4 shows the components and connections.

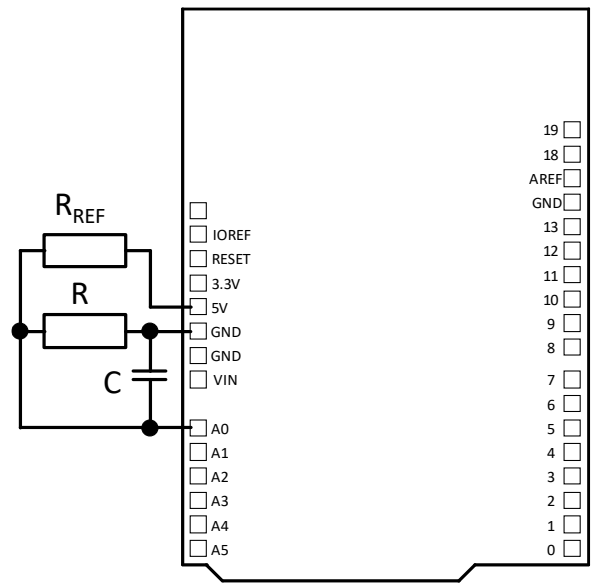

**Figure 4.** The Ohmmeter's components and connections. The 10 nF capacitor at the analogue input of the ADC is needed, it reduces dynamic loading effects and noise.

All four Ohmmeters used $R_{REF}$ of 10 kΩ with a specified tolerance of 0.1 %. We have applied the following code to measure the resistance as an average of ten samples and to send the result to the serial monitor of the Arduino integrated development environment:

```
#define R_REF 10000.0
#define AVERAGES 10     // range: 1 to 100

void setup() {
  analogReference(DEFAULT);
  Serial.begin(9600);
}

void loop()
{
  float resistance;
  uint16_t x;
```

```
    static float cumulativeResistance = 0;
    static uint8_t i = 0;

    x=analogRead(A0);
    resistance = R_REF*x/(1024.0-x);
    cumulativeResistance = cumulativeResistance + resistance;
    i++;
    if (i==AVERAGES)
    {
      Serial.println(cumulativeResistance/AVERAGES);
      cumulativeResistance = 0;
      i = 0;
    }
    delay(500/AVERAGES);  // about two measurements per second
}
```

We have measured the eight resistors with all four Ohmmeters, see table 1 for the results.

| $R$ [Ω] 0.01% | calculated $|\Delta R/R|$ [%] | Arduino board #1 $R$ [Ω] | $|\Delta R/R|$ [%] | Arduino board #2 $R$ [Ω] | $|\Delta R/R|$ [%] | Arduino board #3 $R$ [Ω] | $|\Delta R/R|$ [%] | Arduino board #4 $R$ [Ω] | $|\Delta R/R|$ [%] |
|---|---|---|---|---|---|---|---|---|---|
| 508.2 | 4.34 | 484.29 | 4.7 | 481.06 | 5.34 | 481.06 | 5.34 | 482 | 5.16 |
| 994.82 | 2.47 | 971.82 | 2.31 | 967.12 | 2.78 | 968.3 | 2.67 | 971.83 | 2.31 |
| 1990.3 | 1.51 | 1962.62 | 1.39 | 1961.22 | 1.46 | 1962.62 | 1.39 | 1961.22 | 1.46 |
| 5733.4 | 0.94 | 5705.52 | 0.49 | 5700.71 | 0.57 | 5693.5 | 0.7 | 5688.84 | 0.78 |
| 9940 | 0.88 | 9922.18 | 0.18 | 9887.36 | 0.53 | 9910.57 | 0.3 | 9879.94 | 0.6 |
| 19942.6 | 0.98 | 19941.52 | 0.01 | 19889.19 | 0.27 | 19932.79 | 0.05 | 19941.52 | 0.01 |
| 47038 | 1.45 | 47174.92 | 0.29 | 46920.67 | 0.25 | 46952.45 | 0.18 | 47206.7 | 0.36 |
| 99795 | 2.46 | 100107.52 | 0.31 | 99407.27 | 0.39 | 99404.71 | 0.39 | 100107.52 | 0.31 |

**Table 1.** Measurement results obtained by using four Arduino Ohmmeters.

All boards provided rather good accuracy in a certain range, performed better than some common DMMs. Figure 5 plots the calculated and measured errors.

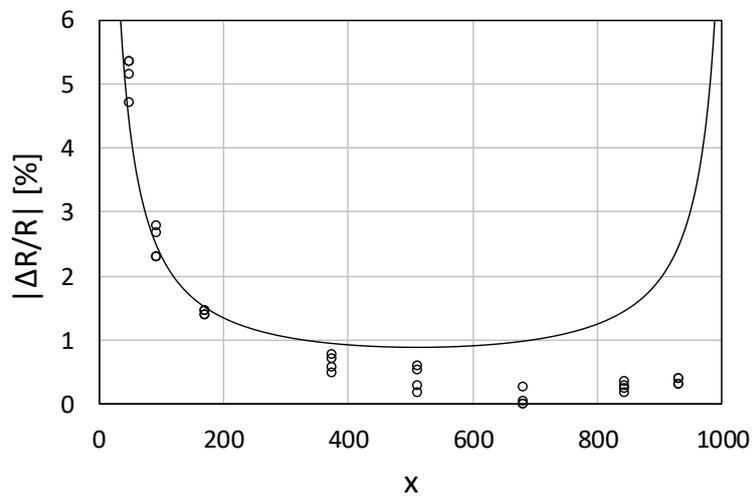

**Figure 5.** Calculated (solid line) and measured (circles) errors listed in table (1) as a function of the ADC code.

The error is considerably lower at higher codes. Note that gain error can produce similar dependence. At codes below 100 the 2-unit absolute error is slightly exceeded. It is normal, since this error specification is only typical, not guaranteed.

Most users are likely to choose easily available 1 % resistors for the reference and test resistors. Additionally, some applications may need different resistance measurement ranges. To demonstrate these situations, we have tested the performance of one Ohmmeter using four different reference resistors. The results are shown in table 2 and in figure 6.

| nominal $R/R_{REF}$ | calculated $|\Delta R/R|$ [%] | $R_{REF}$=1 kΩ, 1 % | | $R_{REF}$=10 kΩ, 1 % | | $R_{REF}$=100 kΩ, 1 % | | $R_{REF}$=1 MΩ, 1 % | |
|---|---|---|---|---|---|---|---|---|---|
| | | R [Ω] | $|\Delta R/R|$ [%] | R [Ω] | $|\Delta R/R|$ [%] | R [Ω] | $|\Delta R/R|$ [%] | R [Ω] | $|\Delta R/R|$ [%] |
| 0.3 | 2.10 | 295.7 | 1.43 | 2978 | 0.72 | 29670 | 1.10 | 297845 | 0.72 |
| 1.0 | 1.78 | 994.2 | 0.58 | 9961 | 0.39 | 99610 | 0.39 | 997271 | 0.27 |
| 3.0 | 2.04 | 3000.0 | 0.00 | 30141 | 0.47 | 303150 | 1.05 | 3001569 | 0.05 |

**Table 2.** Measurement results obtained by using four different reference resistors. The errors represent the deviation from the nominal values, they contain the tolerance of the measured resistors too.

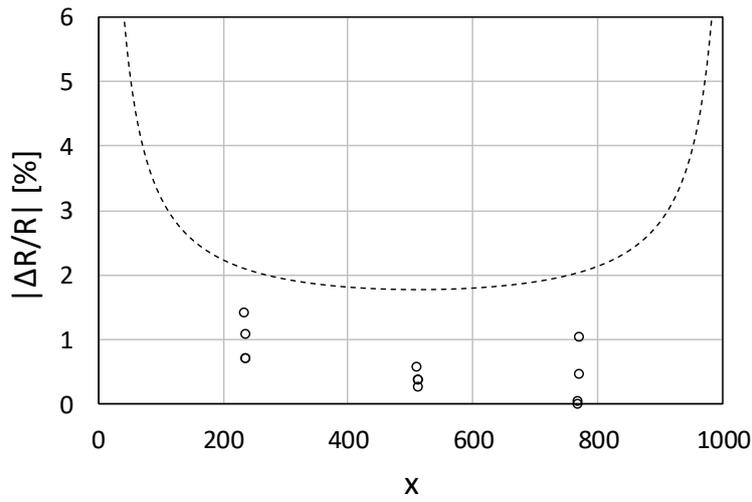

**Figure 6.** Theoretical (dashed line) and measured (circles) errors listed in table (2) as a function of the ADC code.

The measured error falls well below the theoretical value, likely because the actual accuracy of 1 % resistors is pretty much within the margins. The results also suggest that the input current of the ADC had no considerable contribution even at higher resistances. However, it can be significant if more signals are measured by multiplexing or if much higher sample rate is used, since in such cases the ADC input means higher dynamic load. This bears more extensive discussion, therefore, we plan to address this question in a separate paper. Anyway, the 10 nF capacitor at the ADC input is needed.

## On-line calculator

If a desired measurement range from $R_{min}$ to $R_{max}$ is given, the optimal value of $R_{REF}$ should be found. The error is the lowest possible over the range when it is the same at $R_{min}$ and $R_{max}$. In this case the distance of the ADC codes $x_{min}$ and $x_{max}$ at the boundaries from the middle are also the same (see equation (6) and figure 2):

$$\frac{N}{2} - x_{min} = x_{max} - \frac{N}{2} \tag{11}$$

which means

$$N = x_{max} + x_{min} \tag{12}$$

Expressing $x_{min}$ and $x_{max}$ by using $R_{min}$ and $R_{max}$ in equation (3) yields

$$N = \frac{N}{1 + \frac{R_{REF}}{R_{max}}} + \frac{N}{1 + \frac{R_{REF}}{R_{min}}} \tag{13}$$

The solution of this is:

$$R_{REF} = \sqrt{R_{max} \cdot R_{min}} \tag{14}$$

This confirms the same formula used in a previous paper [4].

What is the measurement range for a given $R_{REF}$ and $|\Delta R/R|$? In this case equation (10) should be used to find the range of the ADC codes $x_{min}$ and $x_{max}$, where the error remains under the desired level. The solution is

$$x_{max,min} = \frac{N}{2} \pm \sqrt{\left(\frac{N}{2}\right)^2 - \frac{N \cdot \Delta x}{\left|\frac{\Delta R}{R}\right| - \left|\frac{\Delta R_{REF}}{R_{REF}}\right|}} \tag{15}$$

Using equation (3) we can get the resistance range:

$$R_{max,min} = \frac{R_{REF}}{\frac{N}{x_{max,min}} - 1} \tag{16}$$

We have developed an on-line calculator to support both calculations [10], its user interface elements can be seen in figure 7.

**Figure 7.** Elements of the user interface of the on-line calculator.

## Conclusion

Arduino Ohmmeters can be valuable tools in any physics lab. Many resistive sensors can be used including thermistors, photoresistors, linear displacement sensors and moisture sensors. Students can build their own instruments, they can learn about instrumentation and measurement accuracy which is essential in teaching experimental physics.

We have analysed the performance of four Arduino Ohmmeters and demonstrated their fairly good accuracy. Nevertheless, we have to draw the attention to some other important points.

1. Try to identify all error sources in any measurement and decide whether they have a considerable contribution in your case.
2. Never use the USB supply in any analogue Arduino application. It can compromise the results even in ratiometric measurements.
3. Keep in mind, that the Δ$x$=±2 absolute error of the ADC code is only a typical value, it is not guaranteed. Since you don't use it in critical applications, it is not a problem in most cases.
4. Calibration needs exceptional care, do it only if you are absolutely sure. However, testing the instrument with known values is important and easy.
5. Use equation (10) to estimate the error of your measurement. Don't think that your instrument can provide better accuracy even if some tests suggest this. Figure 8 warns you that overestimating the accuracy may compromise reliability! Learn more about the accuracy of ADCs and William Tell [11].
6. Instrumentation requires special attention and experience. Building, analysing and testing own instruments can help a lot to improve the knowledge in this field. Arduino is a very good platform to support this.

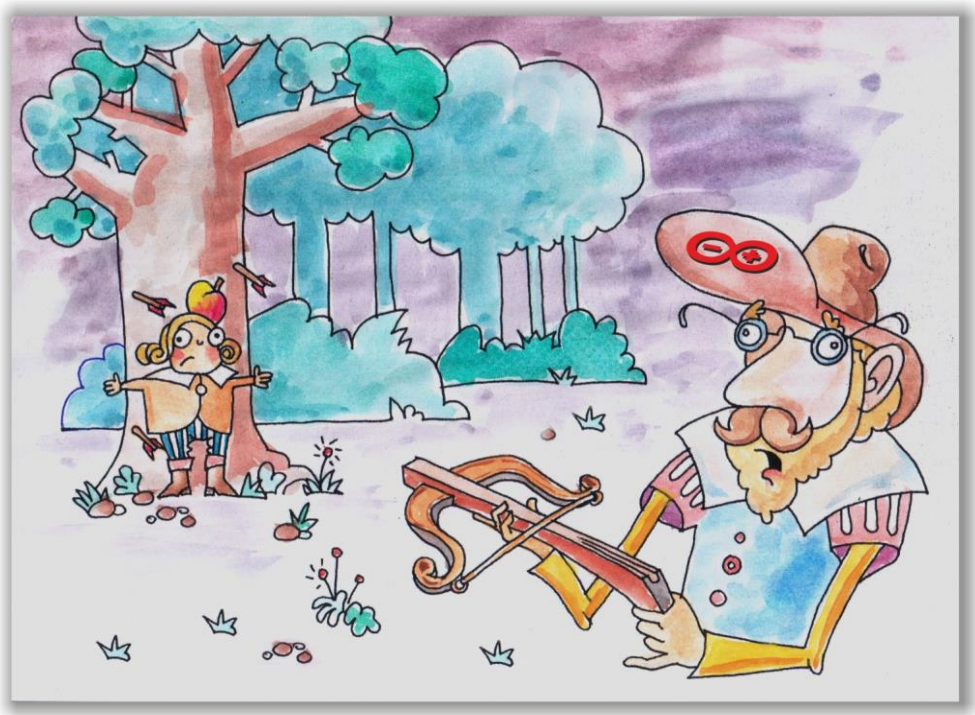

**Figure 8.** Overestimating the accuracy may compromise reliability! [12]

## Invitation

Finally, we invite you to take part in a project! We ask you to build one or more Arduino Ohmmeters and measure 1 kΩ, 3 kΩ, 10 kΩ, 30 kΩ and 100 kΩ resistors using a 10 kΩ reference resistor. All resistors should have 1 % specified tolerance. You can enter your results into an on-line form [10]. This way we can make statistics of the performance of many Arduino Ohmmeters and we'll share the results openly. Thank you for your collaboration in advance, we hope it will be instructive and fun!

## Acknowledgments

This study was funded by the Content Pedagogy Research Program of the Hungarian Academy of Sciences. The authors thank Csaba Magyar for his imaginative illustration.